\documentclass[10pt,conference]{IEEEtran}
\IEEEoverridecommandlockouts
\usepackage[]{collab}
\usepackage{enumitem}
\usepackage{balance}
\usepackage{xcolor}
\usepackage{setspace}
\usepackage{subfigure}
\usepackage{multirow}
\usepackage{multicol}
\usepackage{tabularx}
\usepackage{siunitx}
\usepackage{float}
\usepackage{circledsteps}
\usepackage{color}
\usepackage{hyperref}
\usepackage{diagbox}
\usepackage{caption}
\usepackage{makecell}
\usepackage{colortbl}
\usepackage{tcolorbox}
\usepackage{mdframed}
\usepackage{listings}
\usepackage{booktabs}
\usepackage[utf8]{inputenc}
\usepackage{calligra}
\usepackage[normalem]{ulem}
\usepackage{indentfirst}
\usepackage{url}
\usepackage{soul}
\usepackage{nicematrix}
\usepackage{fontawesome5}
\usepackage{textcomp}
\usepackage{stfloats}
\usepackage{verbatim}
\usepackage{graphicx}
\usepackage{amsmath,amsfonts}
\usepackage[linesnumbered,ruled,vlined,noend]{algorithm2e}
\usepackage{array}
\usepackage[utf8]{inputenc}

\collabAuthor{zh}{purple}{Zhihan Jiang}
\collabAuthor{hy}{magenta}{Haiyu}
\definecolor{ballblue}{rgb}{0.13, 0.67, 0.8}
\collabAuthor{jy}{ballblue}{Jinyang Liu}
\collabAuthor{hx}{teal}{Xiao He}

\usepackage{tikz}

\newcommand\nm{LogPilot\xspace}

\newcommand\company{Volcano Engine\xspace}

\newcommand{\ie}{{\em i.e.},\xspace}
\newcommand{\eg}{{\em e.g.},\xspace}

\newcommand{\boxmargin}{1mm}

\newtcolorbox{myboxa}[2][]{
    colback=gray!10!white,
    colframe=black, enhanced,
    attach boxed title to top left={yshift=-2mm,xshift=5mm},
    title=#2,#1
}
\newtcolorbox{myboxb}[2][]{
    boxsep=3pt,
    left = \boxmargin, right = \boxmargin, top = \boxmargin, bottom = \boxmargin,
    title={#2},#1
}
\newtcolorbox{myboxc}{
    colback=gray!15!white,
    arc = 0pt, outer arc = 0pt,
    boxsep=0pt, left = 3pt, right = 0pt, top = 0pt, bottom = 0pt, 
    leftrule=3pt, bottomrule=0pt,toprule=0pt, rightrule=0pt,
    left = \boxmargin, right = \boxmargin, top = \boxmargin, bottom = \boxmargin
}
\newtcolorbox{myboxd}{
    colback=gray!10,%gray background
    colframe=black,% black frame colour
    width=\columnwidth,% Use 8cm total width,
    arc=1mm, auto outer arc,
    boxrule=0.5pt,
}

\definecolor{myyellow}{HTML}{FFF2CC}
\newcounter{finding}

\definecolor{myyellow}{HTML}{FFF2CC}
\newcounter{insight}

\newcounter{challenge}

\definecolor{mygreen}{HTML}{AFCFA5}
\newcounter{opportunity}

\begin{document}

\thispagestyle{plain}
\pagestyle{plain}

% \title{\nm: an LLM-Powered Framework for Log-Based Alert Diagnosis in Online Service Systems}
\title{\nm: Intent-aware and Scalable Alert Diagnosis for Large-scale Online Service Systems}

\author{
    \IEEEauthorblockN{
        Zhihan Jiang$^{\P}$, 
        Jinyang Liu$^{\ddag}$,
        Yichen Li$^{\ddag}$, 
        Haiyu Huang$^{\P}$,
        Xiao He$^{\ddag}$, \\
        Tieying Zhang$^{\ddag\ast}\thanks{$^{\ast}$Tieying Zhang is the corresponding author.}$,
        Jianjun Chen$^{\ddag}$, 
        Yi Li$^{\ddag}$, 
        Rui Shi$^{\ddag}$, 
        Michael R. Lyu$^{\P}$, 
    }
    \IEEEauthorblockA{
        $^{\P}$The Chinese University of Hong Kong, \{zhjiang22, hyhuang25, lyu\}@cse.cuhk.edu.hk \\
        $^{\ddag}$ByteDance, \{jinyang.liu, liyichen.325, xiao.hx, tieying.zhang, jianjun.chen, liyi.ly, shirui\}@bytedance.com \\
    }
}

\maketitle

\begin{abstract}
Effective alert diagnosis is essential for ensuring the reliability of large-scale online service systems.
However, on-call engineers are often burdened with manually inspecting massive volumes of logs to identify root causes.
While various automated tools have been proposed, they struggle in practice due to alert-agnostic log scoping and the inability to organize complex data effectively for reasoning.
To overcome these limitations, we introduce \nm, an intent-aware and scalable framework powered by Large Language Models (LLMs) for automated log-based alert diagnosis. \nm introduces an intent-aware approach, interpreting the logic in alert definitions (\eg PromQL) to precisely identify causally related logs and requests. To achieve scalability, it reconstructs each request's execution into a spatiotemporal log chain, clusters similar chains to identify recurring execution patterns, and provides representative samples to the LLMs for diagnosis. This clustering-based approach ensures the input is both rich in diagnostic detail and compact enough to fit within the LLM's context window. Evaluated on real-world alerts from Volcano Engine Cloud, \nm improves the usefulness of root cause summarization by 50.34\% and exact localization accuracy by 54.79\% over state-of-the-art methods. With a diagnosis time under one minute and a cost of only \$0.074 per alert, \nm has been successfully deployed in production, offering an automated and practical solution for service alert diagnosis.
\end{abstract}


\section{introduction}

With the rise of cloud computing, many traditional systems have migrated to cloud platforms as large-scale online services. These services support critical business operations globally, making reliability essential. Despite rigorous testing efforts, unexpected failures still occur in production, often causing significant disruptions and financial losses~\cite{aws, google, azure,huang2024faultprofit}.

To ensure service reliability, cloud operators deploy monitoring systems that collect low-level metrics such as per-request latency and response codes. These metrics are aggregated into Service Level Indicators (SLIs), such as request success rate or 99th percentile latency~\cite{frey2013key, chen2022adaptive}. SLIs are then continuously evaluated against predefined Service Level Objectives (SLOs)~\cite{DBLP:conf/icac/DingCSMS19, chen2021graph, li2022intelligent}. When violations occur, alerts are triggered and dispatched to on-call engineers (OCEs) for investigation.

While alerts indicate that something has gone wrong, they often reveal only the symptom of a deeper issue rather than its root cause. To resolve the underlying problem, engineers must perform root cause analysis (RCA). This process demands engineers to examine vast amounts of fine-grained monitoring data.
Among these data, logs, which record detailed runtime events, are the most widely used and crucial diagnostic resource~\cite{DBLP:conf/sigsoft/ZhaoWLPWPWFWZSP21,DBLP:conf/sigsoft/HeZHXLKMWDRL22,huang2024faultprofit,huang2024demystifying}. For instance, one empirical study found that 93\% of failures in distributed systems are manifested in logs, and logs are instrumental in diagnosing nearly 90\% of faults~\cite{DBLP:journals/infsof/YuanLSL20}. This finding aligns with our own practical experience, where logs consistently serve as the primary source of evidence during investigations.

In current industrial practice, this log-based diagnostic procedure typically involves two phases (§~\ref{sec: background_current_practice}): (1) \emph{alert-driven log scoping}, in which engineers filter massive volumes of logs to extract relevant entries; and (2) \emph{investigative log RCA}, which infers the underlying cause of the triggered alert.
This manual process is time-consuming and labor-intensive, motivating a variety of automated log-based RCA approaches from both academia~\cite{DBLP:conf/europar/WittkoppWK24,DBLP:conf/icse/0003W0JCYL25} and industry~\cite{DBLP:conf/sigsoft/Jiang0YC0ZFYYL25,DBLP:conf/sigsoft/ZhangXQHQLZLDLC21}.

However, we find that existing automated approaches fall short in complex industrial scenarios due to fundamental limitations in both phases.
First, for \emph{log scoping} (§\ref{subsec:motivation_log_filtering}), current methods like keyword searches or anomaly detection are alert-agnostic~\cite{DBLP:conf/sigsoft/ZhangXLQZDXYCLC19,DBLP:conf/ijcai/MengLZZPLCZTSZ19,DBLP:conf/icse/YuYFZXWMH24}. Lacking the specific context of the alert, they either overwhelm engineers with irrelevant logs or miss critical evidence entirely.
Second, for \emph{investigative log RCA} (§~\ref{subsec:motivation_log_RCA}), traditional pattern-learning techniques~\cite{DBLP:conf/europar/WittkoppWK24,DBLP:journals/tnsm/NotaroHCG23,huang2024demystifying} are impractical. They rely on large, manually labeled datasets of past failures, which are costly to create and maintain.
While recent LLM-based approaches~\cite{DBLP:conf/eurosys/ChenXMKGSCGFWZG24,DBLP:conf/iclr/XuZZHZLPHZ025,DBLP:conf/icse/0003W0JCYL25} show promise, their practical application is hindered by ineffective log data organization. This either leads to log data volumes exceeding the LLM's context window size or presents interleaved and fragmented event information, thereby impairing the model's reasoning capabilities.

To bridge these gaps, we identify two key opportunities for building a practical and automated alert diagnostic framework. First, we recognize that an intent-aware diagnostic process is achievable by leveraging an alert's definition logic (e.g., Prometheus Query Language (PromQL) expression) to precisely scope the logs relevant to the triggered alert.
Second, we observe that the diagnosis can be made scalable even when the volume of scoped logs is large. This is because the underlying execution paths are not random but follow recurring patterns~\cite{kimura2019proactive,DBLP:conf/sigsoft/ZhangXQHQLZLDLC21}.
We can distill these patterns into a few representative examples, which are both compact enough for an LLM’s context window and rich enough to support accurate, holistic reasoning about the root causes.

Motivated by these insights, we propose \nm, to our knowledge, the first intent-aware and scalable framework powered by LLMs for automated log-based alert diagnosis in large-scale online services.
\nm consists of three key phases.
First, in the \emph{Intent-Aware Log Scoping} phase, an alert-log correlation agent interprets the semantic intent of alerts, typically encoded in PromQL expression, to automatically generate a tailored log filtering tools.
These tools extract the logs and corresponding request IDs that are causally linked to the alert, ensuring high precision in log scoping.
Second, the \emph{Request-Centric Log Chain Processing} phase reconstructs a coherent execution for each request by parsing raw logs into structured events and organizing them into spatiotemporal chains. This representation captures the end-to-end execution flow across distributed components, facilitating deeper diagnostic reasoning.
Finally, in the \emph{Clustering-Based LLM Diagnosis} phase, \nm groups requests based on the similarity of their log event patterns and selects a representative request from each cluster for detailed analysis by a log-based RCA agent.
This approach ensures that the diagnostic input remains within the LLM's context window while retaining essential diagnostic clues.
The identified root causes are then aggregated by a diagnostic summary agent, which synthesizes a final, comprehensive alert diagnosis report.

We conducted an extensive evaluation of \nm on real-world alert data from large-scale production services at Volcano Engine. The results show that \nm significantly outperforms the state-of-the-art baselines~\cite{DBLP:conf/eurosys/ChenXMKGSCGFWZG24,DBLP:conf/iclr/XuZZHZLPHZ025}, improving root cause summarization by 50.34\% (human-evaluated Usefulness) and root cause localization by 54.79\% (Exact Match).
With a response time of under one-minute and an approximate cost of only \$0.074 per alert, \nm proves both practical and scalable.
Furthermore, \nm has been successfully deployed in our production services, and we share our experience and practical insights from this deployment.

In summary, the main contributions of this paper are:
\begin{itemize}[leftmargin=*, topsep=1pt, parsep=0pt]
    \item We investigate the gap between existing log-based diagnosis research and industrial practice, identifying key opportunities for building practical, automated solutions (§\ref{sec:motivation}).
    \item We introduces \nm, the first intent-aware and scalable LLM-based framework that diagnoses service alerts by correlating and analyzing massive volumes of logs (§\ref{sec:method}).
    \item We extensively evaluate and successfully deploy \nm on our production systems, demonstrating its effectiveness, scalability, and practical value (§\ref{sec:evaluation}).
\end{itemize}

\section{Background}

\subsection{Online Service Monitoring and Automated Alerting}
\label{sec: background_monitoring}

Modern large-scale online services are built as complex, distributed systems where reliability is essential to business success.
To uphold this reliability, operators rely on advanced monitoring infrastructure that continuously tracks the health and performance of the system.

\begin{figure}[t]
    \centering
    \includegraphics[width=\columnwidth]{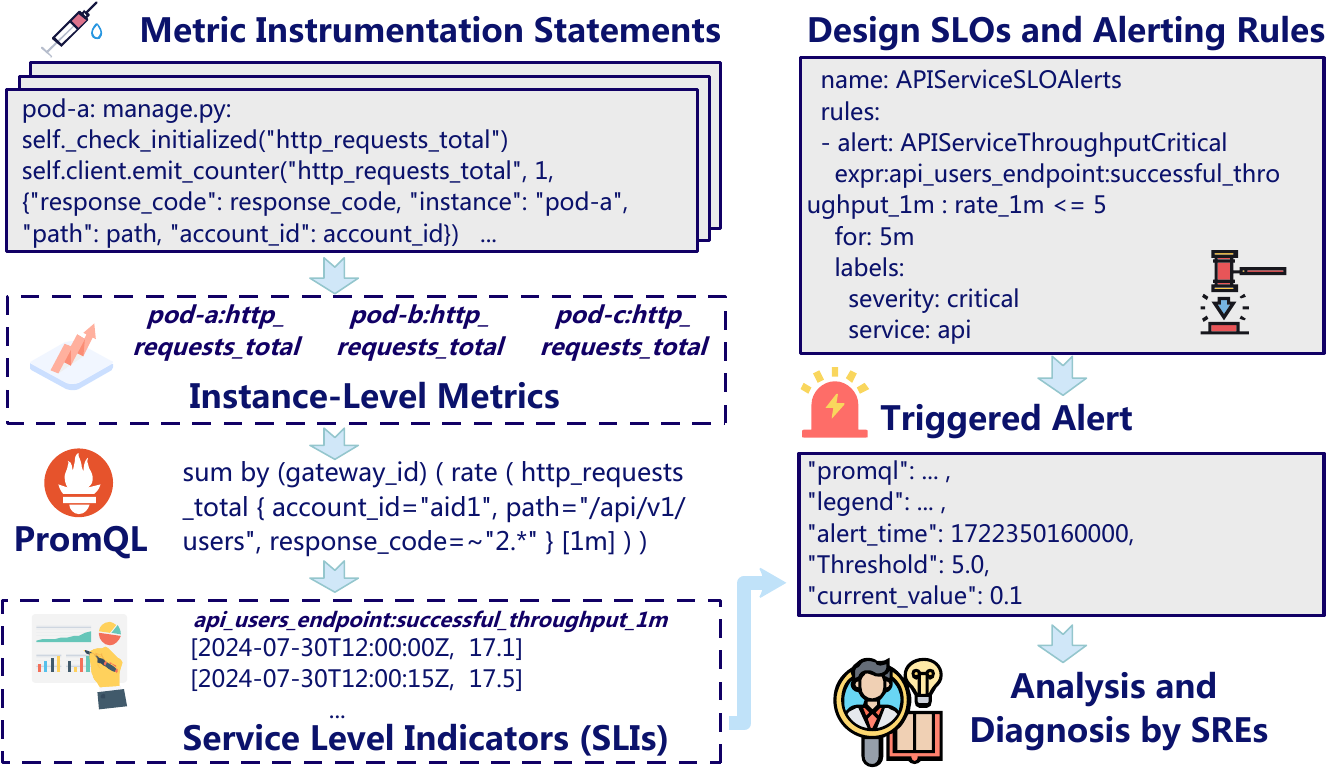}
    \caption{An example service monitoring and alerting.}
    \label{fig: alert_example}
\end{figure}

\noindent
\textbf{Service Metric Collection and Aggregation.}
As shown in Fig.~\ref{fig: alert_example}, gaining visibility into service health begins with instrumentation, where developers embed code to emit performance metrics at runtime.
For instance, a web server might expose a counter like \texttt{http\_requests\_total}, using labels to add dimensions such as the HTTP status code (\texttt{code="403"}) or the endpoint (\texttt{path="/api/v1/users"}).
These low-level metrics are then scraped by a system like Prometheus~\cite{turnbull2018monitoring} and stored in a Time-Series Database (TSDB).

However, these raw metrics from individual instances are too granular to directly reflect overall service health.
A single forbidden request on one server, for example, is mostly considered as noise; its significance emerges only when aggregated with data from all instances to calculate a service-wide error rate.
These aggregated signals constitute the Service Level Indicators (SLIs)~\cite{frey2013key}.
Operators use tools, such as the PromQL~\cite{PromQL-cite}, to perform this aggregation, dynamically computing critical indicators like the request success rate from the raw time-series data according to defined business logic.

\noindent
\textbf{Service Alerting.}
Automated alerting is driven by rules that continuously evaluate SLIs against their corresponding Service Level Objectives (SLOs)~\cite{DBLP:conf/icac/DingCSMS19}. SLOs define the acceptable range for SLIs, expressed as a target threshold (\eg SLI $\leq$ target) or a bounded interval (\eg lower bound $\leq$ SLI $\leq$ upper bound).
These alerting rules are often managed by Prometheus, which evaluates PromQL expressions in real time.
As illustrated in Fig.~\ref{fig: alert_example}, an alert is triggered if a rule's condition is met for a predefined duration, \eg the 1-minute average for the request success rate drops below 95\%.
Once triggered, the rule enters a ``firing'' state and dispatches a notification to OCEs.

\subsection{Current Industrial Alert Diagnosis Practice}
\label{sec: background_current_practice}

While service alerts are effective at detecting anomalous symptoms, they rarely pinpoint the underlying causes. This responsibility consequently falls to OCEs, who perform RCA by manually investigating various observability data. Among these, logs, which provide a detailed, chronological record of system runtime behavior, are the most critical and are typically the first data source engineers utilized during diagnosis in industrial environments~\cite{DBLP:journals/infsof/YuanLSL20,DBLP:conf/sigsoft/ZhangXQHQLZLDLC21}.
This diagnostic process typically involves the following two distinct phases.

\noindent
\textbf{Phase 1: Alert-Driven Log Scoping.}
Diagnosis begins with OCEs forming an initial hypothesis using contextual information from the alert. Alerts typically include key details such as the affected service, the type of failure (\eg low success rate), and the anomalous timeframe.
Based on this context, engineers manually query distributed log storage systems to retrieve logs that are potentially relevant, such as those containing specific error codes or keywords.
This targeted querying helps reduce the search space from billions of log entries to a more manageable subset that is closely related to the alert.

\noindent
\textbf{Phase 2: Investigative Log RCA.}
Following the \emph{log scoping} phase, OCEs proceed to analyze the retrieved subset of logs to infer potential root causes. However, this task is often complicated by the interleaved nature of log entries from numerous service requests, which can obscure the execution flow of any single request. Furthermore, these logs may only capture partial execution statuses, such as error results, without detailing the preceding events that could reveal the root cause.
To obtain a comprehensive picture, OCEs first extract request IDs recorded in the suspicious logs, then use them to trace and search all related log entries across components.
This process relies on request IDs being embedded in the logs, which is automatically handled by the structured logging framework~\cite{structured_logging, logging-google}.
This practice is common across modern industrial systems (\eg Google and WeChat) and widely used frameworks (\eg Spring Cloud), where unique request IDs and other metadata are routinely injected into logs via structured logging~\cite{fse2023nezha}.
These identifiers allow OCEs to correlate all logs associated with a single request as it traverses various system components.
This correlation facilitates the reconstruction of the request’s end-to-end execution path, enabling OCEs to reason about the underlying issues leading up to the alert.
\section{Motivation}
\label{sec:motivation}

The manual, two-phase process for investigating service alerts, while thorough, is slow and labor-intensive.
Although automated solutions exist, they frequently fall short when applied to the complexities of large-scale commercial systems.
This disparity between the practical demands of alert response and the capabilities of current tools motivates our research.
We frame the principal challenges and our identified opportunities within the two core phases of the diagnostic workflow.

\subsection{Alert-Driven Log Scoping}
\label{subsec:motivation_log_filtering}

\noindent
\textbf{Existing Approaches.}
Currently, OCEs rely on two primary methods to isolate relevant logs of an alert. The most common approach involves keyword searches using terms such as \textit{error} or the name of an affected service. More advanced methods use statistical anomaly detection, training models on historical logs to identify deviations from normal patterns~\cite{DBLP:conf/sigsoft/ZhangXLQZDXYCLC19,DBLP:conf/ijcai/MengLZZPLCZTSZ19,DBLP:journals/corr/abs-2107-05908,DBLP:conf/kbse/LeZ21,DBLP:conf/icse/YuYFZXWMH24}. Both approaches aim to extract a small subset of relevant entries from an overwhelming volume of log data.

\noindent
\textbf{The Industrial Gap.}
A fundamental gap in industrial practice lies in the inability to reliably connect an alert's symptom to its underlying log data.
This requires isolating the complete set of logs for each failed request (the ``signal''), from the massive stream of logs generated by successful operations (the ``noise''). In a large-scale system, this signal-to-noise ratio can be exceptionally low.
Existing solutions fall short due to their \textbf{lack of alert intent awareness}.
Whether based on keyword search or anomaly detection, they do not account for the specific semantics of the triggering alert. As a result, they often overwhelm engineers with irrelevant logs (false positives) or miss critical information (false negatives). Crucially, they cannot perform a targeted search for the specific failure condition defined by the alert.

\noindent
\textbf{Opportunity.}
\label{sec: log_scoping_opportunity}
We identify a critical and underexplored asset: the semantic intent embedded in the alert itself.
The alert is defined by a PromQL expression (see Sec.~\ref{sec: background_monitoring}), which encodes key metrics, labels, and aggregation logic.
We observe that these contextual cues in PromQL often align well with the log data generated by our services.
Our internal analysis further shows that over \textbf{96.4\%} of alerts include label fields (\eg account ID, gateway, error code, pod name, and endpoint), whose meanings are reflected in the logs, even if the exact field names differ.
This underscores a valuable opportunity: Rather than relying on static rules or loose keyword matching, we can use the semantic signals encoded in the alert to guide precise log extraction.
By automatically interpreting the alert's intent, our approach would precisely extract the necessary logs, creating a focused and complete diagnostic picture.

\subsection{Investigative Log RCA}
\label{subsec:motivation_log_RCA}

\noindent
\textbf{Existing Approaches.}
Once relevant logs are scoped, the next step is RCA. Traditional RCA approaches based on pattern learning~\cite{DBLP:conf/europar/WittkoppWK24,DBLP:journals/tnsm/NotaroHCG23} often require large labeled datasets and generalize poorly to evolving or unseen failures. Recently, LLMs have shown promise due to their strong reasoning capabilities and ability to operate without task-specific training, offering a more adaptable solution.

\noindent
\textbf{The Industrial Gap.}
Despite their strengths, LLMs encounter two significant gaps in real-world industrial applications.
The first is \textbf{scale and complexity}: In large-scale online services, a single failure alert may involve hundreds of log lines per request, and often spans thousands of failed requests.
Existing approaches~\cite{DBLP:conf/eurosys/ChenXMKGSCGFWZG24,DBLP:conf/iclr/XuZZHZLPHZ025,DBLP:conf/icse/0003W0JCYL25} lack effective strategies for organizing this vast and heterogeneous log data. As a result, the input to LLMs either exceeds their context window or consists of fragmented, interleaved events across requests, both of which undermine the model’s ability to perform accurate and comprehensive reasoning.
The second gap is the implicit assumption that logs contain all necessary information for a diagnosis.
In practice, logs can be incomplete, inconsistently formatted, or entirely missing critical events~\cite{DBLP:journals/infsof/YuanLSL20,DBLP:conf/iwpc/BogatinovskiNA022}. This assumption limits the effectiveness of existing approaches when the logs are insufficient for root cause identification.
% \hx{We also need to discuss the gap for the work of recent LLM-agent for RCA on cloud incidents \cite{DBLP:conf/eurosys/ChenXMKGSCGFWZG24}.}

\noindent
\textbf{Opportunity.}
Although raw logs are extensive and varied, the logs of specific request types often reveal consistent execution patterns, especially for requests that fail due to the same underlying root cause~\cite{DBLP:conf/sigsoft/ZhangXQHQLZLDLC21,DBLP:conf/sigsoft/Jiang0YC0ZFYYL25}. 
By identifying these patterns, we can distill the raw logs of each request into compact and informative execution patterns.
Selecting representative examples for each pattern offers compact diagnostic clues that well-suited to an LLM’s context window, enabling more accurate and holistic reasoning.
Furthermore, even if these patterns prove insufficient for diagnosis, LLMs can surface such limitations directly in the analysis. This provides actionable feedback to improve service observability.

\section{Methodology}
\label{sec:method}

\begin{figure*}[t]
    \centering
    \includegraphics[width=\linewidth]{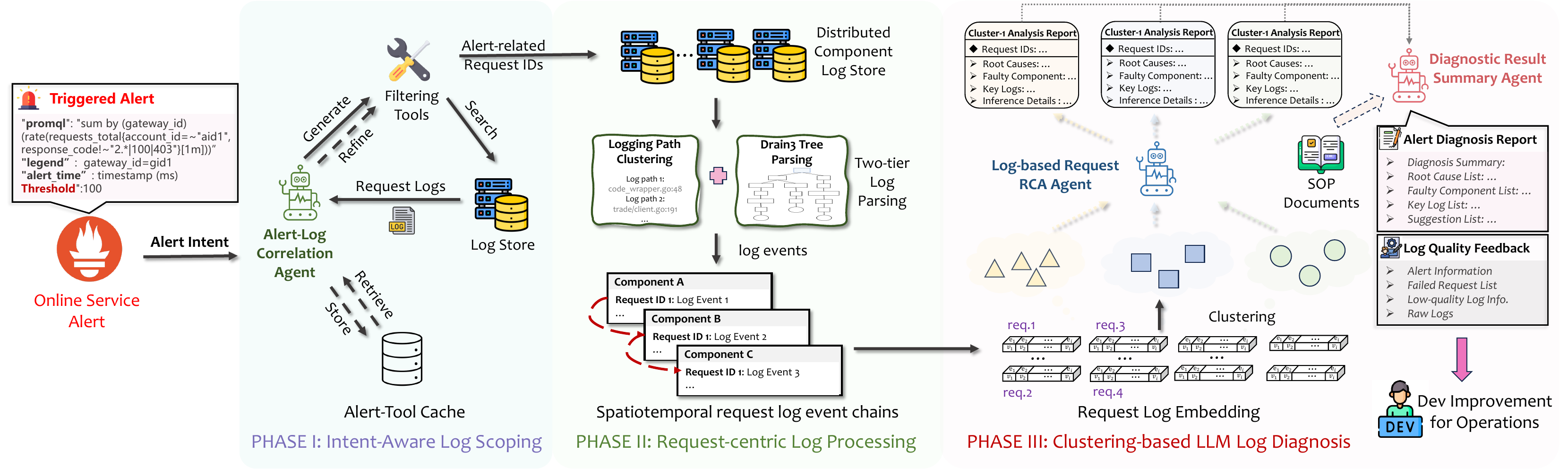}
    \caption{The overall framework of \nm.}
    \label{fig:method_framework}
\end{figure*}

\subsection{Overview}

To bridge the above industrial gaps, we propose \nm, an intent-aware and scalable log-based framework that leverages LLMs for automated and scalable alert diagnosis in large-scale online services. The overall framework of \nm, depicted in Fig.~\ref{fig:method_framework}, operates in three sequential phases.

The first phase, \emph{Intent-Aware Log Scoping}, is triggered upon alert activation. Here, an \emph{alert-log correlation (ALC) agent} interprets the semantic intent behind the alert, typically encoded as PromQL expressions, and automatically generates customized, lightweight and executable log filtering tools for each alert. These tools retrieve logs and extract request IDs that are causally linked to the alert.

In the second phase, \emph{Request-Centric Log Chain Processing}, \nm reconstructs detailed log event chains for each relevant request. This is accomplished by first parsing raw logs into structured log events using a two-tier parsing module. The parsed events are then chronologically and spatially organized into log chains that preserve inter-component execution flows, thereby enabling comprehensive request-level analysis.

Finally, the \emph{Clustering-Based LLM Diagnosis} phase, focuses on efficient and accurate RCA. To recognize the common execution pattern of alert-related requests, \nm clusters them based on the similarity of their log chain embeddings. A representative request from each cluster is selected and analyzed by an \emph{LLM-based request RCA agent} to identify underlying issues. The resulting insights are synthesized by a \emph{diagnostic result summary agent}, which produces a comprehensive alert diagnosis report.

\subsection{Intent-aware Log Scoping}

Once an alert is triggered, the first step is to collect and filter logs that are related to the investigation.
As highlighted in Sec.~\ref{subsec:motivation_log_filtering}, conventional methods that rely on static rules or anomaly detectors often produce an overwhelming volume of irrelevant logs or, conversely, omit crucial information.

\begin{figure}[t]
    \centering
    \includegraphics[width=\columnwidth]{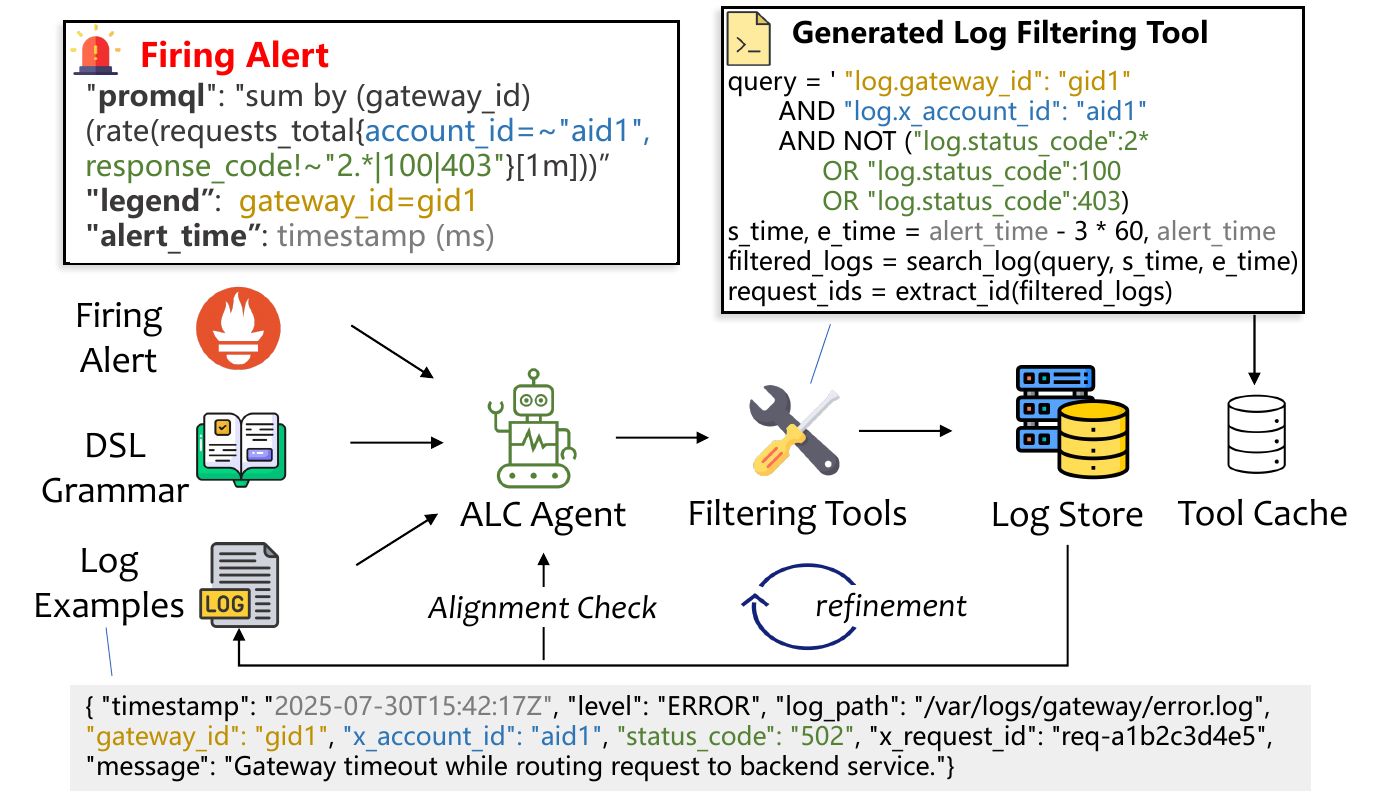}
    \caption{An example workflow of \emph{alert-log correlation} agent}
    \label{fig: log_scoping}
\end{figure}

To overcome these limitations, \nm introduces an \emph{ALC agent} that interprets the semantic intent behind an alert and extracts log data with high precision.
The key insight is that the logs from service requests that directly triggered the alert are more informative for analysis.
This causal linkage can be inferred from the alert’s definition logic, most commonly expressed in PromQL, as introduced in Sec.\ref{sec: background_monitoring}.
For instance, Fig.~\ref{fig: log_scoping} illustrates an alert that triggers when the one-minute growth rate of requests with specific \textit{gateway\_id}, \textit{account\_id} and response code surpasses a threshold.
As explored in Sec.\ref{sec: log_scoping_opportunity}, such contextual information can guide the correlation between alerts and logs.
However, due to diverse logging schemas across different services, naive schema-based matching is infeasible (see Fig.~\ref{fig: log_scoping}).
To overcome this obstacle, we leverage LLMs to heuristically interpret and align the semantic content of alerts and logs.
However, directly using LLMs to understand alert intent and process large volumes of raw log messages is computationally expensive.
To address this, we instead harness their strong code generation capabilities~\cite{DBLP:journals/corr/abs-2406-00515,DBLP:journals/pacmse/Mu00YZWL024,DBLP:conf/icse/Du0WWL0FS0L24,DBLP:conf/icse/DiLYJCCCCCCFGGH24,zhong2025ccisolver} to automatically generate lightweight, customized log filtering tools tailored to each alert.

\noindent
\textbf{1. Tool Generation.}
The ALC agent is prompted to generate a runnable log filtering tool using three key inputs, as shown in Fig.~\ref{fig: log_scoping}: (1) \emph{Alert definition}: The detailed definition and generation logic of the alert; (2) \emph{Log Examples}: A random sampling of logs to illustrate the specific service schema. (3) \emph{Domain Specific Language (DSL) Grammar}: The documentation details the grammar and usage of our log query language.
By integrating these inputs, the agent generates a tool (\ie a Python script with DSL query) that retrieves logs from the relevant time window, matching the exact criteria specified in the alert. 
This tool then extracts the request IDs causally linked to the alert from these logs.

An example of such a generated tool is shown in Fig.~\ref{fig: log_scoping}.
This LLM-based approach enables dynamic adaptation to heterogeneous alert definitions and service log schema, eliminating the need for manually writing or maintaining filtering logic across a large and evolving alert surface.

\noindent
\textbf{2. Tool Refinement.}
Following the generation of a log scoping tool, it is used to query the log store to extract key logs and request IDs.
However, one-shot LLM tool generation can be unreliable due to the inherent stochasticity of LLMs~\cite{DBLP:conf/acl/Bi0W0GLZS0S24,DBLP:conf/forge/PengGLXSS25}.
Consequently, we introduce a feedback-based iterative refinement process to ensure the high quality of the generated tools.
Specifically, both the query results and any execution errors are treated as feedback.
This feedback enables the agent to perform an alignment check: it verifies whether the retrieved logs conform to the alert logic.
If a mismatch occurs, the agent refines the tools accordingly.
The refinement loop repeats until alignment is achieved or the maximum number of iterations (set to 3) is reached.
In rare cases (less than 3\% in deployment) where no valid tools emerge, human experts intervene to create tools for subsequent use and caching.

\noindent
\textbf{3. Tool Caching.}
Once a tailored log filter tool is validated, it can be cached for future reuse. Specifically, if a new alert with an identical PromQL definition is triggered, the corresponding tool can be retrieved from the cache and applied directly, bypassing the need for regeneration. This caching mechanism reduces LLM invocation overhead and enhances response latency, especially in high-frequency alerting environments.

\subsection{Request-centric Log Processing}

In real-world production systems, log data from service requests are often interleaved and scattered across distributed component log stores, which complicates analysis. To address this, we organize this complex data in a request-centric manner, mirroring the process of manual diagnosis.

\noindent
\textbf{1. Two-tier Log Parsing.}
After filtering for alert-related service requests, we use the request IDs to concurrently collect all raw logs for each request from distributed service components. Raw logs are typically unstructured and substantial, impeding automated analysis. Consequently, log parsing has been widely adopted as a critical prerequisite to convert them into structured log events~\cite{DBLP:journals/csur/HeHCYSL21,DBLP:conf/sigsoft/HeZHXLKMWDRL22,DBLP:conf/issta/JiangL00HGCZL24,jiang2024lilac,huang2025no}.

In production systems, we observe that in most log sources, logging paths with the exact code lines are contained within the logs.
However, logs with the same logging path can still record different log events due to the dynamic propagation and construction of logging variables~\cite{DBLP:conf/kbse/HuoLSHXL23,li2024go,li2024exploring,DBLP:conf/icse/0003W0JCYL25,huang2025no, zhu2019tools, zhu2023loghub}.
Therefore, we employ a two-tier log parsing approach.
First, we extract the logging path from each raw log and perform a coarse-grained clustering based on this path. Log messages without a path are grouped together.
Then, within each coarse-grained group, we use the Drain~\cite{DBLP:conf/icws/HeZZL17} algorithm to heuristically parse the raw logs into structured log events.

\noindent
\textbf{2. Request Log Event Chain Construction.}
To support effective reasoning by LLMs about system behavior, it is critical to structure log data in a manner that captures the interaction patterns among system components. To this end, we construct a spatiotemporal log event chain for each request, leveraging log timestamps and component identifiers. This representation preserves both the chronological execution flow and the invocation relationships across components.

Formally, let the log event list for a single request be denoted as $L = \{e_1, e_2, \cdots \}$. To reduce the volume of log data and ensure it remains within the context window of modern LLMs, we deduplicate this list by retaining only the first occurrence of each unique log event and then sort the result chronologically: $L' = \{e_1’, e_2’, \cdots \}$. We then partition the sorted list $L'$ into segments based on the associated service components, yielding a chain structure that spans multiple components: $C_1 = \{e_1’, e_2’, \cdots, e_i’\}$, $C_2 = \{e_{i+1}’, e_{i+2}’, \cdots\}$, and so on. The ordering of these component segments is determined by the timestamp of their earliest log event, thereby preserving the actual invocation sequence across services.

\subsection{Clustering-based LLM Diagnosis}

After processing the logs for each request into an event chain, the next step is RCA. However, diagnosing all requests is impractical due to the cost, inefficiency, as well as the limited LLM context issue.
Based on the observation that requests with similar log event patterns often stem from the same underlying issue. This insight allows for a significant reduction in the diagnostic workload, as a single alert can typically be traced to a small number of root causes~\cite{DBLP:conf/sigsoft/ZhangXQHQLZLDLC21,liu2023scalable,DBLP:conf/icse/LiuHCLKZHZLXRZL23,DBLP:conf/iclr/XuZZHZLPHZ025}.
Consequently, \nm focuses on analyzing representative request for each distinct execution pattern.

\nm achieves this by employing a clustering-based approach, which consists of several stages.
First, \nm vectorizes the log event chain of each request to create a log pattern embedding.
Next, it groups these requests into clusters based on embedding similarity. Then, within each cluster, a \emph{log-based request RCA agent} analyzes one representative request to identify the root cause. Finally, the \emph{diagnostic result summary agent} aggregates the results from all clusters to produce a final comprehensive diagnosis report for the alert.

\noindent
\textbf{1. Request Log Embedding.}
To quantitatively represent each request's pattern for similarity measurement, we transform the log data into a vector embedding following previous work~\cite{DBLP:conf/usenix/LouFYXL10,DBLP:conf/icse/0002SL24,DBLP:conf/kbse/ZhangJLNCMSP24}.
Specifically, the embedding for a single request is represented as a vector  $V = [v_1, v_2, \cdots, v_n]$, where $n$ represents the total number of distinct log events across all requests, and $v_i$ denotes the count of the $i$-th log event. This representation captures both the presence and frequency of log events, providing a holistic view of request log patterns.

To assess the similarity between these embeddings, we use a log-scaled cosine similarity metric.
This approach mitigates the issue where high-frequency events can disproportionately dominate the similarity score, thereby better capturing the underlying structural log patterns. 
First, for any request embedding vector $V = [v_1, v_2, \cdots, v_n]$, we apply a logarithmic scaling transformation to produce a new vector $V' = [v'_1, v'_2, \cdots, v'_n]$, where each element is calculated as: $v'_i = \log(1 + v_i)$.
% $$
% v'_i = \log(1 + v_i)
% $$
Subsequently, the similarity between two requests, represented by their scaled vectors $V'_A$ and $V'_B$, is computed using the standard cosine similarity formula:
\begin{small}
\begin{equation}
\small
    \text{Similarity}(V'_A, V'_B) = \frac{V'_A \cdot V'_B}{\|V'_A\| \|V'_B\|} % = \frac{\sum_{i=1}^{n} v'_{A,i} v'_{B,i}}{\sqrt{\sum_{i=1}^{n} v'_{A,i}^2} \sqrt{\sum_{i=1}^{n} v'_{B,i}^2}}
\end{equation}
\end{small}
This ensures that requests are represented based on their fundamental log patterns, making the process less susceptible to variations in system load.

\noindent
\textbf{2. Request Clustering.}
To identify distinct groups of alert-related issues, we employ the Hierarchical Agglomerative Clustering (HAC) algorithm on the request log embeddings.
The algorithm initializes each request as a separate cluster and progressively merges the most similar pairs until their similarity score falls below a predefined threshold.
We chose this algorithm because it does not require predefining the number of clusters.
The resulting clusters represent groups of requests with similar log patterns, from which we can infer a shared underlying root cause.
Based on our experience, the number of resulting clusters is typically small (under 10), as a specific alert usually has only a few root causes~\cite{DBLP:conf/icse/LiuHCLKZHZLXRZL23,DBLP:conf/ccgrid/YuOPWCSJWLP24}.

\begin{figure}[t]
    \centering
    \includegraphics[width=\columnwidth]{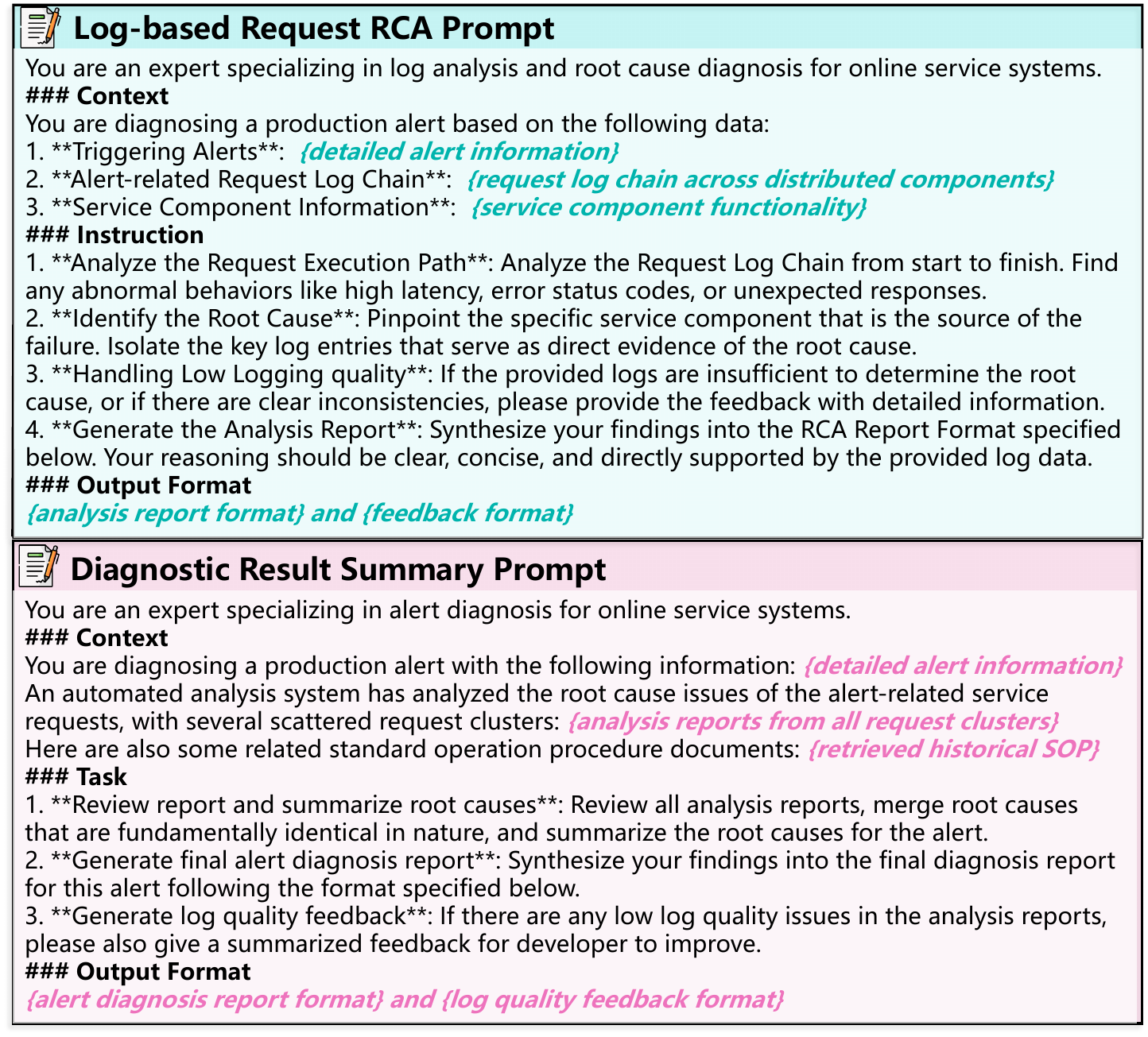}
    \caption{Example prompt templates for \emph{log-based request RCA} and \emph{diagnostic result summary} agent.}
    \label{fig: prompt_example}
\end{figure}

\noindent
\textbf{3. Log-based Request RCA Agent}
This LLM agent performs RCA for each alert-related request cluster with a common execution pattern.
Since analyzing every request in a large cluster is impractical, our framework mimics manual troubleshooting by selecting a single, representative request for analysis.
To identify this request from a cluster of $m$ requests, we first calculate the average log embedding vector (the centroid): $\bar{V} = \frac{1}{m} \sum_{i=1}^m V_i$.
We then select the request whose log embedding is closest to this centroid, $\bar{V}$, as representative request for further detailed analysis.

Leveraging the analytical and reasoning capabilities of LLMs, the \emph{Log-based Request RCA Agent} analyzes the selected request's lifecycle to identify the root cause of the issue.
As shown in the example prompt in Fig.~\ref{fig: prompt_example}, the agent takes the detailed alert information, the request's log chain, and service component information as input. It then generates an analysis report that includes the identified root causes, the faulty component, key log evidence, and inference details. Furthermore, considering that logging quality can vary, the agent is designed to handle cases where logs are insufficient or inconsistent to determine the root cause (\eg error information is not transmitted or logged). In such situations, it identifies and reports this poor logging quality in its analysis.

\noindent
\textbf{4. Diagnostic Result Summary Agent}
The RCA for each cluster is performed independently and in parallel. Consequently, the findings from different clusters can be interrelated or contain overlapping information (\eg they may stem from the same root cause but exhibit different symptoms).
To address this, \nm incorporates a \emph{Diagnostic Result Summary Agent} to aggregate these disparate results.
As illustrated in Fig.~\ref{fig: prompt_example}, this agent takes alert information and all individual analysis reports as input, and synthesizes them into a final, comprehensive alert diagnostic report, including a summary, a unified list of root causes, faulty components, and key logs. 

Furthermore, to provide domain-specific insights and actionable recommendations, we use a knowledge base of historical Standard Operating Procedure (SOP) documents. Employing the Retrieval-Augmented Generation (RAG) paradigm, the agent retrieves the most relevant SOPs based on the identified root causes of all request clusters. This context is then supplied to an LLM to generate a list of suggestions for resolving the alert. Additionally, the agent identifies instances of poor logging quality into a feedback report for developers, highlighting opportunities for improving future DevOps processes. More detailed cases can be found in Sec.~\ref{sec: deployment_experience}. 

\section{Evaluation Setup}
\label{sec:evaluation}

We first evaluate \nm by answering the following research questions (RQs):
\begin{itemize}[leftmargin=*, topsep=1pt, parsep=0pt]
    \item \textbf{RQ1:} How effective is \nm in alert diagnosis?
    \item \textbf{RQ2:} How robust are the components of \nm?
    \item \textbf{RQ3:} What is the efficiency and cost of \nm?
\end{itemize}
We also share our deployment experience of \nm in the \company Cloud systems (§~\ref{sec: deployment_experience}).

\subsection{Dataset Collection}
To assess the effectiveness of our proposed method, we collected alert data from four large-scale online services at our company, each serving millions of users globally. The data collection period spanned from June 15 to July 15, 2025.
From this initial set, we randomly sampled a subset of alerts that can be manually analyzed. %by Site Reliability Engineers (SREs).
This resulted in a final dataset of 202 alerts, with detailed statistics presented in Table \ref{tab: dataset_statistics}.

\subsection{Dataset Labeling}
The complete log data corresponding to each alert's timeframe is archived in our internal log storage, and a large portion of the selected alerts were manually diagnosed by engineers at the time of their occurrence and formally documented in failure review reports.
The manual annotation of our dataset was conducted by a team of four annotators. This team comprised two Ph.D. students with over two years of research experience in system operations, and two industry engineers, each with more than five years of experience in software development and maintenance.
For each alert, following previous work~\cite{DBLP:conf/icse/0003W0JCYL25}, the annotators carefully reviewed the available diagnostic data and labeled from two aspects:
\begin{itemize}[leftmargin=*, topsep=1pt, parsep=0pt]
    \item \textbf{Root Cause Summarization}: A concise description of the root cause was formulated through a collaborative discussion among the annotators. The primary objective of this summary is to provide a clear and actionable guide for system maintainers to diagnose similar issues in the future.
    \item \textbf{Root Cause Localization}: The specific components of the service identified as the root cause were pinpointed. These components are also reflected in the fields of the log messages. While most cases were attributed to a single root cause component, some alerts involved two or three components being identified as the source of the issue.
\end{itemize}

\subsection{Evaluation Metrics}

\begin{itemize}[leftmargin=*, topsep=1pt, parsep=0pt]
    \item \textbf{Root Cause Summarization:}
    To evaluate the quality of root cause summary, following prior research~\cite{DBLP:conf/icse/Wang00GW21, DBLP:conf/icse/SahaH22, DBLP:journals/pacmse/KangAY24, DBLP:conf/icse/0003W0JCYL25}, we use the following metrics: (1) \textit{Automated Evaluation}: \emph{METEOR}~\cite{DBLP:conf/acl/BanerjeeL05} and \emph{ROUGE}~\cite{lin2004rouge} are employed to measure the textual similarity between generated summaries and the ground truth. To capture contextual meaning beyond lexical overlap, we also computed \emph{Semantic Similarity} by embedding the summaries using the \textit{Doubao-1.5-Embedding} model~\cite{doubao-embedding}. (2) \textit{Human Evaluation}: We assessed the \emph{Usefulness} of the summaries, defined as how accurately they explain the cause of an issue.
    Evaluations were conducted by highly experienced production engineers.
    For each issue, evaluators reviewed both the ground truth and the corresponding summaries from all methods, assigning a usefulness score ranging from 0 to 1. The final score for each baseline is averaged across all evaluators.
    \item \textbf{Root Cause Localization:} We measure localization accuracy using: (1) \textit{Exact Match}: This metric measures the percentage of instances where the predicted set of root cause components is exactly identical to the ground-truth set. (2) \textit{Top-3 Accuracy}: This metric assesses whether the set of ground-truth components is fully contained within the top-3 most likely components proposed by the model.
\end{itemize}

\begin{table}[]
\footnotesize
\centering
\caption{Evaluated alerts and unique root causes (URCs)}\label{tab: dataset_statistics}
\setlength{\tabcolsep}{4pt}
\begin{tabular}{cccccc}
\toprule
 Datasets &  Service $\mathcal{A}$ & Service $\mathcal{B}$  & Service $\mathcal{C}$ & Service $\mathcal{D}$ &  Total \\ \midrule
\# Alerts  & 35  & 40  &  52 &  75 &  202 \\ \midrule
\# URCs  & 27  &  29 & 43 &  62 & 161 \\ \midrule
\end{tabular}
\end{table}

\subsection{Baselines}
We exclude traditional log diagnosis methods~\cite{DBLP:conf/ccs/Du0ZS17,DBLP:conf/sigsoft/ZhangXLQZDXYCLC19,DBLP:conf/ijcai/MengLZZPLCZTSZ19,DBLP:conf/kbse/LeZ21} due to their reliance on labeled training data and limited explainability, two critical drawbacks that make them unsuitable for production systems.
Recently, many studies~\cite{DBLP:conf/sigsoft/ZhangGBWM0R24,DBLP:conf/sigsoft/RoyZBBLFR24,DBLP:conf/naacl/HuangZZ25} have explored the use of LLM for RCA in online services. We compare \nm against the following state-of-the-art methods with the same LLM backend:
\begin{itemize}[leftmargin=*, topsep=1pt, parsep=0pt]
    \item \textbf{LLM with sampling}~\cite{DBLP:conf/iclr/XuZZHZLPHZ025}: This method directly feeds downsampled log data to an LLM for RCA.
    \item \textbf{RCA Agent}~\cite{DBLP:conf/iclr/XuZZHZLPHZ025}: A multi-agent system that enables an LLM to write and execute code to process large volumes of log data, identifying root causes from the processed results.
    \item \textbf{RCACopilot}~\cite{DBLP:conf/eurosys/ChenXMKGSCGFWZG24}: This approach uses alert information, corresponding log data, and retrieved historical SOPs to perform RCA. To accommodate the LLM's context window, we only input logs with a \textit{warning} or \textit{error} level.
\end{itemize}

\subsection{Implementation Details}
\nm is implemented as an internal service that integrates with our alert system.
It leverages an internal log store search engine to query and filter log data from distributed services components.
The similarity threshold for request clustering is empirically set to 0.7, with its performance impact evaluated in RQ2.
For the LLM backend selection, \nm is powered by the latest \textit{Doubao} series on the Volcano Engine platform through the official API provided.
Specifically, the \emph{alert-log correlation} and \emph{log-based request RCA} agents utilize the \textit{Doubao-Seed-1.6-thinking} model to leverage its strong reasoning capabilities, while the \emph{diagnostic result summary} agent defaults to the \textit{Doubao-Seed-1.6-flash} model which prioritizes speed for lightweight tasks.

\section{Evaluation Results}

\subsection{RQ1: Effectiveness of Alert Diagnosis}

The evaluation results of both root cause summarization and localization dimensions are presented in Tab.~\ref{tab:rq1-result}.

\noindent
\textbf{Root Cause Summarization.}
In the task of root cause summarization, \nm demonstrates remarkable performance improvements over existing baselines across automated and human-evaluated metrics. Regarding automated metrics, compared to the strongest baseline, RCA Agent, \nm achieves average gains of 20.75\% in ROUGE-1, 27.37\% in METEOR, and 8.66\% in semantic similarity. These results demonstrate the high quality of the summaries generated by \nm.
Furthermore, in our human evaluation, the Usefulness metric directly measures the practical benefit of a summary to system maintainers.
In this crucial assessment, \nm obtains an average improvement of 50.34\% over RCACopilot, underscoring its ability to produce not just accurate, but also clear and actionable diagnostic information.

\noindent
\textbf{Root Cause Localization}
\nm's performance in root cause localization also achieves state-of-the-art, where it markedly surpasses all baseline methods in identifying the faulty components associated with an alert.
The best-performing baseline, RCA Agent, is consistently outperformed by \nm across all systems and metrics.
On average, \nm delivers a 54.79\% higher score in Exact Match and a 22.49\% higher score in Top-3 Accuracy. The most significant performance gap is observed in the Top-3 Accuracy for Service A, where \nm's score is 26.92\% greater than RCA Agent's. These findings highlight the precision of our approach in pinpointing the faulty components.

\begin{table*}[t]
    \centering
    \footnotesize
    \caption{Root cause analysis results from both ~\textit{summarization} and ~\textit{localization} dimensions for each service.}
    \label{tab:rq1-result}
        \begin{tabular}{l|l||cccc|cc}
            \toprule
               \multirow{2}{*}{\textbf{System}} & \multirow{2}{*}{\textbf{Model}} &
                \multicolumn{4}{c}{\textbf{Root Cause Summarization}} &
                \multicolumn{2}{c}{\textbf{Root Cause Localization}} \\
            \cmidrule{3-8}
                & &
                \textbf{ROUGE-1} &
                 \textbf{METEOR}  &
                \textbf{Semantics} &
                \textbf{Usefulness} &
                \textbf{Exact Match} &
                \textbf{Top-3 Acc.} \\
            \midrule
               \multirow{4}{*}{Service $\mathcal{A}$} &  LLM w/ BS &  0.381 & 0.303 & 0.713 & 0.454 & 0.343 &  0.543 \\
                & RCACopilot &  0.468 & 0.374 & 0.795 & 0.563 & 0.429 &  0.686 \\
                & RCA Agent &  0.475 & 0.369 & 0.812 & 0.640 & 0.457 &  0.743 \\
                & \nm  &  \textbf{0.587} & \textbf{0.522} & \textbf{0.894} & \textbf{0.823} & \textbf{0.743} &  \textbf{0.943} \\
            \midrule
               \multirow{4}{*}{Service $\mathcal{B}$} &  LLM w/ BS &  0.375 & 0.322 & 0.719 & 0.473 & 0.400 &  0.475 \\
                & RCACopilot &  0.483 & 0.397 & 0.804 & 0.570 & 0.450 &  0.725 \\
                & RCA Agent &  0.512 & 0.408 & 0.831 & 0.668 & 0.475 &  0.775 \\
                & \nm &  \textbf{0.602} & \textbf{0.497} & \textbf{0.913} & \textbf{0.851} & \textbf{0.725} &  \textbf{0.950} \\
            \midrule
               \multirow{4}{*}{Service $\mathcal{C}$} &  LLM w/ BS &  0.352 & 0.309 & 0.688 & 0.427 & 0.346 &  0.481 \\
                & RCACopilot &  0.413 & 0.346 & 0.755 & 0.494 & 0.423 &  0.673 \\
                & RCA Agent &  0.432 & 0.367 & 0.808 & 0.588 & 0.442 &  0.747 \\
                & \nm &  \textbf{0.538} & \textbf{0.460} & \textbf{0.852} & \textbf{0.785} & \textbf{0.712} &  \textbf{0.885} \\
            \midrule
               \multirow{4}{*}{Service $\mathcal{D}$} &  LLM w/ BS &  0.401 & 0.356 & 0.748 & 0.492 & 0.387 &  0.533 \\
                & RCACopilot &  0.517 & 0.437 & 0.821 & 0.609 & 0.507 &  0.733 \\
                & RCA Agent &  0.538 & 0.446 & 0.846 & 0.643 & 0.560 &  0.787 \\
                & \nm &  \textbf{0.631} & \textbf{0.539} & \textbf{0.924} & \textbf{0.895} & \textbf{0.800} &  \textbf{0.960} \\
            \bottomrule
            \end{tabular}%
\end{table*}

\subsection{RQ2: Robustness of Individual Components}

\noindent
\textbf{Robustness of Intent-aware Log Scoping.}
To evaluate the robustness of the log scoping process, we randomly sample 60 alerts from our datasets and manually assess the quality of the generated log filter tools. Each tool is assigned a score of 0 if it fails to execute or returns empty results (\eg due to error conditions). Otherwise, a human annotator rate its quality on a scale from 0 to 1, based on inspection of both its logic and execution results.
We compared the original alert-log correlation agent with two ablation settings: \nm \emph{w/o log examples} and \nm \emph{w/o tool refinement}.

\begin{figure}[t]
    \centering
    \includegraphics[width=\columnwidth]{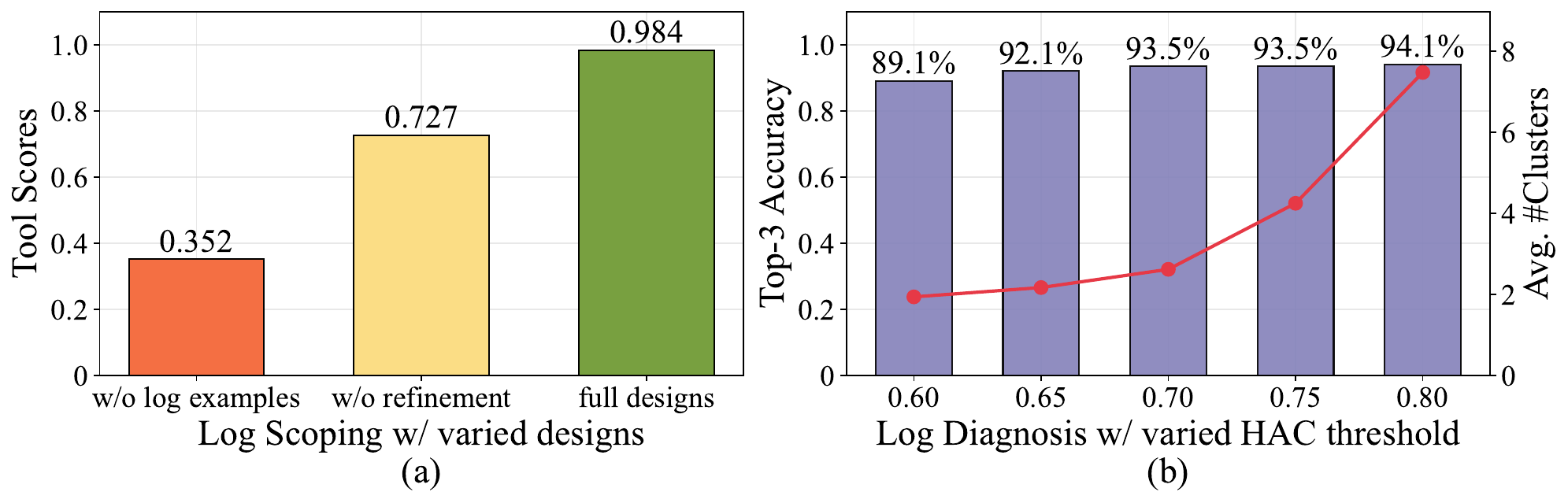}
    \caption{Robustness evaluation of different phases in \nm.}
    \label{fig: RQ2_robustness}
\end{figure}

Fig.~\ref{fig: RQ2_robustness}-(a) summarizes the results: (1) Without illustrative log examples for the LLMs, the quality of the generated tool is significantly diminished and largely unusable. This is primarily due to a lack of knowledge regarding the specific service's log schema. (2) In the absence of the feedback-based tool refinement procedure, the agent is prone to generating unreliable tools, for instance, by mismatching the log schema with PromQL conditions. This results in an average quality score of only 0.727. (3) With all designs integrated, our \nm  achieves a high average score of 0.984 in the log filtering tool quality assessment, demonstrating its effectiveness in correlating alerts with logs by generating executable tools.

\noindent
\textbf{Robustness of Clustering-based LLM Diagnosis.}
The clustering step is critical for accurate LLM-based diagnosis and is controlled by a manually set threshold $\theta_{\text{HAC}}$. We performed a sensitivity analysis to examine its impact, measuring both the system’s top-3 RCA accuracy and the average number of request clusters as $\theta_{\text{HAC}}$ varies.

As shown in Fig.~\ref{fig: RQ2_robustness}-(b):
(1) The top-3 accuracy of \nm remains stable and high (around 94\%) for $\theta_{\text{HAC}}$ values ranging from 0.60 to 0.80.
(2) Increasing $\theta_{\text{HAC}}$ produces more fine-grained clusters, leading to a greater number of analyzed requests and potentially more LLM invocations. Thanks to the aggregation capabilities of our summarization agent, overall RCA performance is maintained.
(3) Reducing $\theta_{\text{HAC}}$ results in fewer clusters and a potential loss of valuable diagnostic information, slightly lowering accuracy (\eg to 89.1\% when $\theta_{\text{HAC}}$=0.60).
Based on this trade-off between diagnostic cost and performance, we select $\theta_{\text{HAC}}$ = 0.7 as the default threshold.

\begin{figure}[t]
    \centering
    \includegraphics[width=\columnwidth]{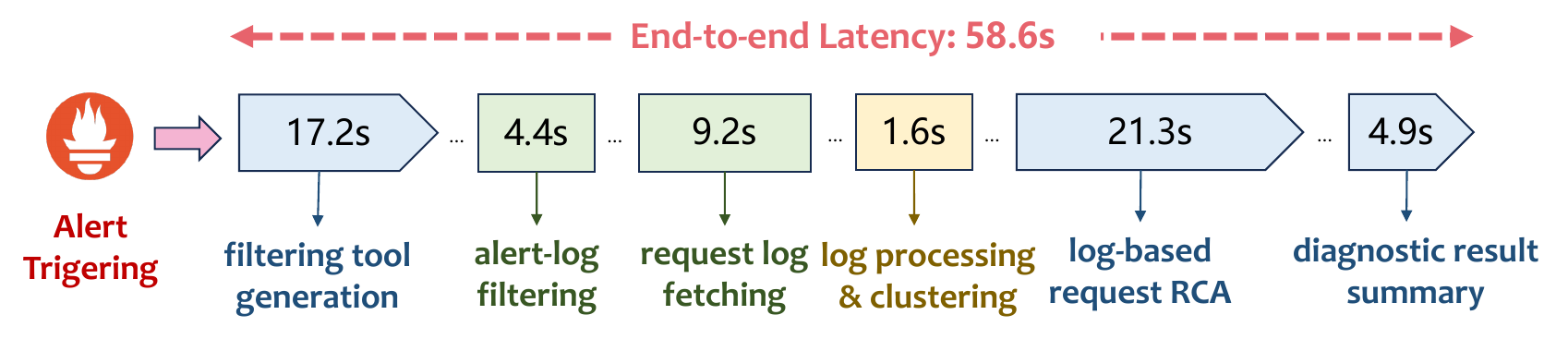}
    \caption{Average E2E and stage-wise latency of \nm.}
    \label{fig: efficiency_time_distribution}
\end{figure}

\subsection{RQ3: Efficiency and Cost of \nm}

In production systems, the high volume of alerts and the demand for timely responses make RCA efficiency critical.
To evaluate \nm in this context, we scale the number of alerts to 548.
Notably, this assessment does not require ground-truth root causes, allowing us to focus solely on efficiency and cost.

\noindent
\textbf{Efficiency.}
We evaluate efficiency by measuring the average processing time across the different stages of \nm, with a breakdown shown in Fig.~\ref{fig: efficiency_time_distribution}.
Our results show that LLM calls dominate the overall runtime. Specifically, generating the log filtering tool and executing the log-based RCA procedure take an average of 17.2s and 21.3s, respectively. The final summarization step, using a streamlined model with shorter prompts, completes in just 4.9s on average.
Despite this reliance on LLMs, the total end-to-end latency remains modest at 58.6 seconds per alert, staying well below the one-minute mark. This represents a substantial improvement over manual diagnosis, which can take several minutes to tens of minutes. Moreover, caching the log filtering tool for recurring alert types can further reduce latency in future diagnoses.

\begin{figure}[t]
    \centering
    \includegraphics[width=\columnwidth]{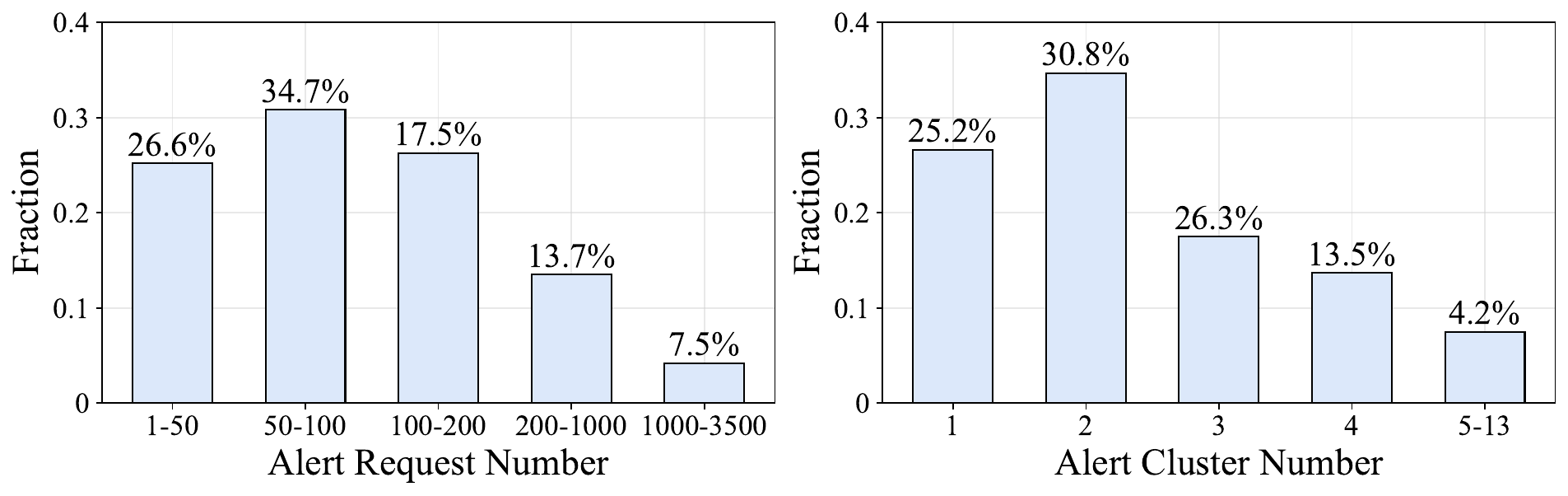}
    \caption{The distribution of request and cluster number.}
    \label{fig: efficiency_cluster_distribtuion}
\end{figure}

\noindent
\textbf{Cost.}
To evaluate economic viability, we first analyze the number of filtered requests and resulting request clusters per alert (Fig.~\ref{fig: efficiency_cluster_distribtuion}).
Although a single alert may associated with over 3,000 requests (mean: 198.65), clustering reduces the number of required LLM invocations to a maximum of 13 (mean: 2.56), achieving a 98.71\% reduction.
Furthermore, token usage statistics show that, per alert, \nm consumes an average of 69.73K prompt tokens and 8.08K response tokens. According to current pricing~\cite{doubao-pricing}, this results in an estimated cost of \$0.074 per alert for diagnosis. Given the critical importance of rapid and accurate alert handling in production systems, we consider this cost both acceptable and justified.

\section{Deployment Experience}
\label{sec: deployment_experience}

\begin{figure}[t]
    \centering
    \includegraphics[width=\columnwidth]{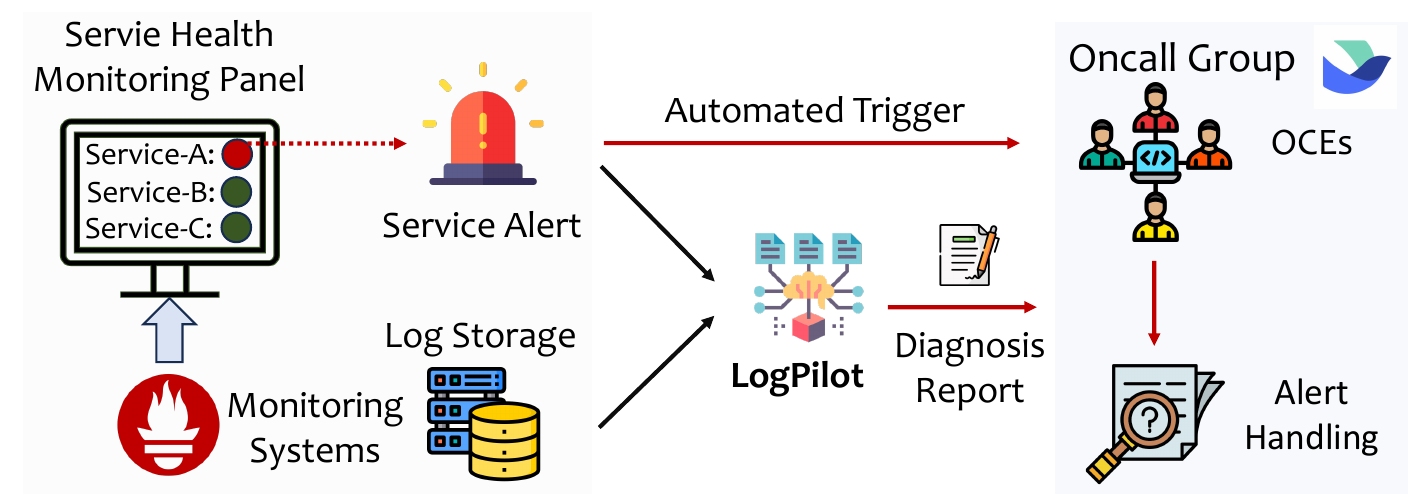}
    \caption{The production deployment workflow of \nm.}
    \label{fig: deployment_workflow}
\end{figure}

\noindent
\textbf{Performance in Production.}
Since June 2025, \nm has been deployed across 12 production services within the Volcano Engine Cloud systems.
Fig.~\ref{fig: deployment_workflow} shows how we have integrated LogPilot with existing service monitoring systems. Upon an alert trigger, a dedicated on-call group is automatically created via Lark to notify related OCEs. 
Simultaneously, \nm autonomously initiates its three-stage RCA process, generating a diagnostic report with actionable insights to assist OCEs in alert triage and handling.
By July 31, 2025, \nm had analyzed over 3,500 production alerts and provided RCA results to assist OCEs.
The practical effectiveness of \nm is demonstrated by user feedback from system architects and engineers. 
The overall acceptance rate for the generated reports is \textbf{84.21\%}. Of these, 60.53\% of the RCA results were identified completely correct (an exact match with the ground truth) and 23.68\% were partially correct (we identified a subset of the ground-truth root causes).
Even in the remaining 15.79\% of cases, the inference details of \nm were still reported to offer valuable insights for subsequent manual investigation.
These statistics highlight \nm’s practical effectiveness in production environments.

\begin{figure}[t]
    \centering
    \includegraphics[width=0.9\columnwidth]{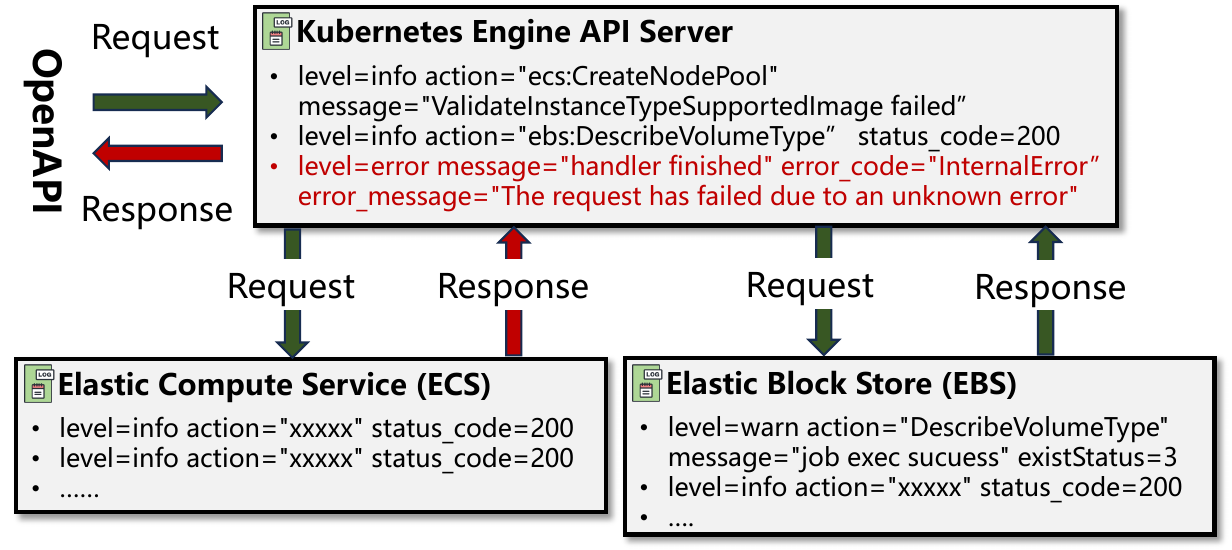}
    \caption{A real-world example of inconsistent logging case.}
    \label{fig: log_quality_case}
\end{figure}

\noindent
\textbf{Enhancing Log Quality and Bridging Dev-Ops.}
Beyond alert diagnosis, \nm provides feedback on log quality, helping to identify observability gaps and recurring logging anti-patterns for developers. A common scenario identified is the ``silent failure'', where a request returns an error code, yet the corresponding logs across all traversed components are recorded at the info level with no indication of abnormality. Such logging practices severely hinder manual diagnosis.
Furthermore, \nm detects log inconsistencies within the request execution lifecycle.
For instance, as depicted in Fig.~\ref{fig: log_quality_case}, \nm identified a case where the cloud control plane's request to create a node pool failed due to an ECS image type validation error. This specific error, however, was not propagated to the originating service and was instead logged as a \textit{info} level by the Kubernetes Engine API Server.
Concurrently, the EBS component, despite returning a success code, anomalously logged at the \textit{warn} level with exit code 3.
\nm successfully flagged these discrepancies and surfaced potential reliability and observability issues that merit further developer attention.
This automated feedback loop represents a step toward identifying software reliability issues and bridging the gap between Development (Dev) and Operations (Ops).
To formalize this process, we have incorporated log quality into the evaluation of engineering discipline across development teams from diverse services. 
Over time, this encourages more consistent and comprehensive logging practices, ultimately improving service maintainability.

\noindent
\textbf{Enriching Alert Context and Future Work.}
We believe \nm, as an automated framework that integrates high-level alerts with low-level logs, fundamentally enriches alerts with deeper system context and analytical insights. This capability paves the way for improving alert management in large-scale online service systems.
Specifically, for a given alert type, the system state at each trigger can vary considerably, even if the triggering rules are identical. By augmenting alerts with the identified root causes derived from real-time underlying log data, we can provide more granular contextual information, \eg the actual root causes. This, in turn, facilitates more precise alert aggregation and the development of more fine-grained alert-handling SOPs. This area constitutes a primary focus for our future research and development efforts.

\section{Discussion}

\subsection{Lessons Learned}

\noindent
\textbf{From Definitive Conclusions to Actionable Insights.}
Prevailing studies in RCA~\cite{DBLP:conf/iwqos/Li0JZWZWJYWCZNS21,DBLP:conf/europar/WittkoppWK24,DBLP:conf/iclr/XuZZHZLPHZ025} traditionally focus on delivering a single, definitive conclusion. However, our practical experience in industrial production environments indicates that the inherent complexity of these systems precludes any single RCA method from achieving complete reliability.
This limitation often hinders the trust of OCEs for automated diagnoses.
Consequently, we advocate for a paradigm shift: moving beyond sole root cause verdicts to the provisioning of actionable diagnostic insights.
We argue that the ability to profile system behavior and distill critical clues from vast telemetry data is of paramount importance.
Our experience and feedback from \nm's users confirm this: the failure clues and inference details \nm provides enable OCEs to rapidly comprehend the system's state and derive actionable insights. The advent of LLMs presents a valuable opportunity to further develop and scale this paradigm.

\noindent
\textbf{From Isolated Signals to Correlated Observability Data.}
While \nm effectively bridges the gap between high-level alerts and underlying runtime logs, modern online services generate a diverse source of observability and failure management data, including metrics, traces, oncall records, and so on.
These data sources often exist in silos. The dynamic correlation and fusion of these heterogeneous signals are crucial for comprehensive system diagnosis and operations, yet this remains a persistent challenge in the industry.
Addressing this gap effectively necessitates the co-design of both data infrastructure and analytical algorithms.
We believe the strong pattern recognition and reasoning capabilities of LLMs open promising new research directions toward achieving unified and intelligent system observability.

\subsection{Generalizability}

Our proposed \nm, is designed for broad applicability across diverse online service systems, offering insights to both the research and practitioner communities.
First, \nm is founded on widely adopted observability data sources: alert information (\eg PromQL definitions~\cite{PromQL-cite}), and fundamental runtime logs.
Given that Prometheus has become the de facto standard for monitoring in modern cloud systems~\cite{turnbull2018monitoring}, and runtime logs represent the most basic source of observability data, \nm can be readily integrated into various online services.
Second, \nm is architected around request-centric analysis, a methodology that reconstructs log event chains at the request level and employs clustering techniques to extract meaningful runtime patterns. 
This methodology is broadly applicable, as request-centric observability remains a core paradigm in diagnosing complex online services~\cite{DBLP:conf/nsdi/SambasivanZRKWSWXG10,DBLP:journals/tsc/ZhouCWZL18,DBLP:conf/iwqos/Li0JZWZWJYWCZNS21}.
Third, the generalization capabilities (\eg in-context learning ability) of LLMs~\cite{DBLP:conf/acl/ZengLLWLD024,DBLP:conf/sigsoft/ZhangGBWM0R24} further enhance \nm's adaptability.
By incorporating system-specific documentation into the agent’s retrieval-augmented context, \nm can be tailored to new systems with minimal human effort.
Finally, while our current deployment of \nm utilizes our internal Doubao-series LLMs due to privacy considerations, prior work has shown that various LLMs offer strong comprehension and reasoning capabilities~\cite{DBLP:conf/eurosys/ChenXMKGSCGFWZG24,DBLP:journals/corr/abs-2502-12521}. Consequently, \nm can be flexibly deployed with various open-source or commercial LLMs, facilitating its broader adoption in industrial settings.

\section{Related Work}

\noindent
\textbf{Log-based Diagnosis.}
Alert or failure diagnosis using logs is vital for ensuring the reliability of software systems. It generally involves two stages: log scoping and RCA.
Log scoping reduces the vast volume of logs to a relevant subset for analysis. Existing approaches~\cite{DBLP:conf/sigsoft/ZhangXLQZDXYCLC19,DBLP:conf/ijcai/MengLZZPLCZTSZ19,DBLP:journals/corr/abs-2107-05908,DBLP:conf/kbse/LeZ21,DBLP:conf/icse/YuYFZXWMH24} often apply anomaly detection to identify logs that deviate from normal patterns. For example, LogRobust~\cite{DBLP:conf/sigsoft/ZhangXLQZDXYCLC19} uses an attention-based Bi-LSTM to detect anomalies, while LogAnomaly~\cite{DBLP:conf/ijcai/MengLZZPLCZTSZ19} leverages a template2Vec embedding to capture log semantics. However, these methods typically ignore the context provided by alerts, limiting their ability to scope logs aligned with specific alert conditions.
The second stage, log-based RCA, identifies the underlying issue of an alert. Prior work~\cite{DBLP:conf/europar/WittkoppWK24,DBLP:journals/tnsm/NotaroHCG23} commonly relies on statistical correlation or pattern mining. For instance, FDiag~\cite{fdiag} uses statistical metrics to associate anomalies with potential root causes, and LogRCA~\cite{DBLP:journals/tnsm/NotaroHCG23} adopts a semi-supervised approach for anomaly localization. Nonetheless, these techniques often struggle with the scale and complexity of log data in production environments.

\noindent
\textbf{Log Analysis and RCA in the LLM Era.}
Recent advances in LLMs have spurred efforts to enhance log analysis and RCA. Several works leverage LLMs’ semantic understanding to interpret log data.
For example, LogGPT~\cite{han2023loggptloganomalydetection} learns normal log patterns using a GPT-based model and applies reinforcement learning (RL) to improve detection. 
Knowlog~\cite{icse24knowlog} integrates domain knowledge into pretrained LMs to enhance log comprehension.
Other studies fine-tune LLMs for domain-specific RCA tasks. OASIS~\cite{assess} adapts GPT-3 to summarize cloud incidents, while ThinkFL~\cite{zhang2025thinkflselfrefiningfailurelocalization} employs RL to equip lightweight LLMs with reasoning abilities.
To better utilize observability data, some approaches adopt retrieval-augmented generation (RAG) or agent-based architectures. RCACopilot~\cite{DBLP:conf/eurosys/ChenXMKGSCGFWZG24} aggregates multi-source data for LLM-based RCA, and RCA-agent~\cite{DBLP:conf/iclr/XuZZHZLPHZ025} generates diagnostic tools using LLMs.
Despite promising results, these methods often struggle to structure the substantial diagnostic data effectively, leading to context window overflows and fragmented or interleaved inputs, which limits their scalability in real-world systems.

\section{Conclusion}
This paper presented \nm, the first intent-aware and scalable LLM-based framework for automated alert diagnosis via logs in large-scale cloud service systems. We identified key limitations in existing automated solutions, including imprecise, alert-agnostic log scoping and the inability to effectively organize massive volumes of log data for reasoning.
To address these, \nm introduces two core innovations. First, it is intent-aware: it leverages the semantics of alert definitions to precisely extract causally related logs. Second, it is scalable: it reconstructs user requests into log chains, clusters them to uncover common execution patterns, and feeds compact, representative samples to the LLM for efficient diagnosis.
Evaluations on production data show that \nm outperforms state-of-the-art baselines in both root cause summarization and localization. With diagnoses under a minute at \$0.074, and successful production deployment, \nm offers a practical solution for enhancing service reliability.

\section{Acknowledgment}

This work was supported by the Research Grants Council of the Hong Kong Special Administrative Region, China (No. SRFS2425-4S03 of the Senior Research Fellow Scheme and No. CUHK 14209124 of the General Research Fund).

\balance
\normalem
\bibliographystyle{IEEEtran}
\bibliography{ASE25}

\end{document}